# Cartographie de l'habitat de reproduction du tétras-lyre (*Lyrurus tetrix*) dans les Alpes françaises

Projet « Tétras-lyre »
SYNTHESE DES RESULTATS
Février – Novembre 2023

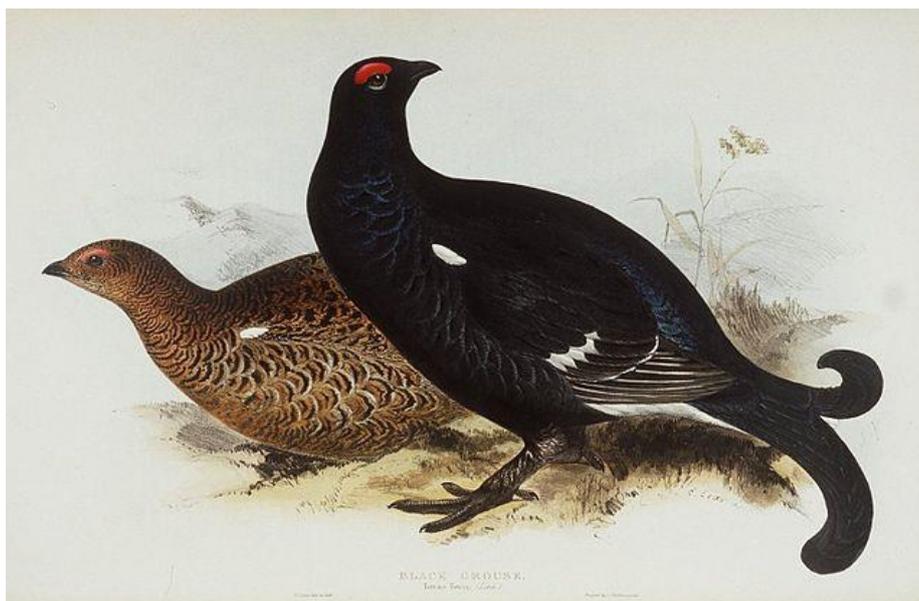

*Rédigé par*

Alexandre Defossez (INRAE)

*Relu par*

Samuel Alleaume (INRAE)

Sandra Luque (INRAE)

Marc Montadert (OGM)

# Table des matières

# I) Introduction

Le tétras-lyre (*Lyrurus tetrix*) est un oiseau galliforme emblématique des Alpes. La protection de son habitat constitue un enjeu de conservation important dans le contexte des changements globaux, et notamment en raison du changement climatique, dont l'impact se manifeste plus intensément dans les milieux montagnards (Knight, 2022). Sur le territoire national, le tétras-lyre fait l'objet de programmes de conservation. Il est essentiellement menacé par la fermeture et la dégradation des milieux ainsi que par les infrastructures et activités humaines liées à l'exploitation du milieu alpin, dont la pratique des sports de montagne.

Le caractère favorable d'un habitat pour le tétras-lyre est d'abord lié à la disponibilité de ses ressources alimentaires. Le type de nourriture consommé varie selon la saison : le tétras-lyre se nourrit en hiver essentiellement de bourgeons et d'épines de pin, tandis qu'en été son alimentation se concentre sur les baies, fleurs et jeunes pousses associées aux landes à éricacées. Nichant en période estivale, les landes à éricacées constituent un habitat-clé pour sa reproduction.

Ce travail prend la suite des analyses réalisées par Josselin Giffard-Carlet en 2022 dans le cadre du même projet. Ces analyses ont notamment permis de clarifier la méthodologie pour produire les différentes variables environnementales susceptibles de prédire l'habitat de reproduction du tétras-lyre. En particulier, les travaux précédemment réalisés ont montré le potentiel d'une classification des strates de végétation à partir d'images à Très Haute Résolution Spatiale (THRS).

Le travail présenté ici s'inscrit dans un partenariat entre l'Institut National de Recherche pour l'Agriculture, l'Alimentation et l'Environnement (INRAE) et l'Office Français de la Biodiversité (OFB). Il vise à proposer une cartographie de l'habitat de reproduction du tétras-lyre à l'échelle des Alpes françaises. Notre approche s'appuie principalement sur le traitement de données issues de la télédétection et en particulier l'imagerie optique satellitaire à différentes résolutions spatiales. Nos travaux ont été réalisés en collaboration étroite avec Samuel Alleaume et Sandra Luque (INRAE) ainsi que Marc Montadert de l'Observatoire des Galliformes de Montagne (OGM).

## II) Objectifs et stratégie méthodologique

Nous chercherons à répondre aux objectifs suivants :

- **Améliorer la connaissance des conditions environnementales favorables à la reproduction du tétras-lyre.**
- **Cartographier l'habitat de reproduction potentiel à l'échelle des Alpes française à partir de l'imagerie satellitaire.**

Nous faisons l'hypothèse que l'habitat de reproduction du tétras-lyre peut être approché à travers la prédiction de la probabilité de présence des nichées. Notre approche méthodologique consiste réaliser un modèle de distribution d'espèce ou *Species Distribution Model* (SDM) appliqué aux nichées du tétras-lyre comme un proxy de son habitat de reproduction potentiel à l'échelle des Alpes françaises. Le principe d'un SDM consiste à modéliser la probabilité de présence d'une ou plusieurs espèces en fonction de données spatialisées pouvant être reliées à diverses caractéristiques de l'environnement. Nous parlerons de « variables environnementales » pour désigner les variables permettant d'expliquer la probabilité de présence des nichées.

La première étape de notre démarche a consisté à produire un ensemble de variables importantes pour la reproduction du tétras-lyre, ou de manière plus générale, capables d'influencer positivement ou négativement la distribution des nichées. Nous avons repris les variables précédemment identifiés lors des précédents travaux. Ces variables environnementales appartiennent à deux groupes : les variables de structure du paysage et les variables phénologiques. Nous faisons l'hypothèse que la distribution des nichées est influencée principalement par des éléments structurant le paysage, notamment la présence de landes ligneuses à éricacées qui fournissent les ressources alimentaires, mais aussi par certaines caractéristiques écologiques de la végétation, comme les variations spatio-temporelles de la productivité (variations phénologiques).

Quatre types de variables environnementales ont été intégrées au SDM :

- *Le Dynamic Habitat Index (DHI) correspondant à une série d'indices phénologiques.*
- *Une cartographie d'occupation du sol (classification) comprenant différentes strates de végétation.*
- *L'hétérogénéité du paysage à travers un indice de texture.*

    Ces variables peuvent être obtenues à partir de l'imagerie satellitaire à Haute (Sentinel-2) et Très Haute Résolution (SPOT6-7).

- *L'altitude.*

    Elle provient des Modèles Numériques de Terrain (MNT) produits par l'Institut National de l'Information Géographique et Forestière (IGN).

Quant aux données d'observation des nichées, elles nous ont été fournies par l'Observatoire des Galliformes de Montagne (OGM).

Les principales étapes de notre démarche sont résumées dans la **Figure 1**.

*Figure 1.* Diagramme de flux représentant les principales étapes du cadre méthodologique.

## III) Matériel

### 1. Imagerie satellitaire
#### a) Images multispectrales

Les images Sentinel-2 (S2), acquises dans le cadre du programme Copernicus (ESA), sont particulièrement adaptées au suivi de la végétation. Elles possèdent à la fois une haute résolution spatiale avec les bandes 2, 3, 4 et 8 à 10 mètres, et une importante résolution spectrale avec 13 bandes spectrales. Les caractéristiques physiques des bandes spectrales utilisées dans le cadre de notre travail sont décrites dans le **Tableau 1**. Les images S2 possèdent enfin une répétitivité temporelle élevée avec environ une acquisition prise tous les 5 jours sur une zone donnée. Cette caractéristique permet de produire des séries temporelles denses et de suivre de manière fine les variations phénologiques de la végétation. Les images S2 sont acquises sur des fauchées de 110 km. Les images S2 ont été téléchargées lors des précédents travaux à l'aide du programme TELIS développé par Benjamin Commandre (https://gitlab.irstea.fr/benjamin.commandre/telis) sur le catalogue Theia (catalogue.theia-land.fr).

*Tableau 1. Description des caractéristiques physiques des bandes Sentinel-2. Source : ESA.*

| Spatial Resolution | Spectral Region | Band | Sentinel-2A satellite | | Sentinel-2B satellite | |
|---|---|---|---|---|---|---|
| | | | Wavelength (nm) | Bandwidth (nm) | Wavelength (nm) | Bandwidth (nm) |
| **10m** | Red | 4 | 664.5 | 38 | 665 | 39 |
| | NIR | 8 | 835.1 | 145 | 833 | 133 |
| | *Narrow* NIR | 8a | 864.8 | 33 | 864 | 32 |
| **20m** | SWIR | 11 | 1613.7 | 143 | 1610.4 | 141 |
| | SWIR | 12 | 2202.4 | 242 | 2185.7 | 238 |

#### b) Images à Très Haute Résolution Spatiales (THRS)

Les images SPOT6 et SPOT7 possèdent une très haute résolution spatiale à 1,5 mètres pour la bande panchromatique et une résolution à 6m concernant les bandes Rouge, Verte, Bleu (RVB) et Proche Infrarouge (PIR). Les images SPOT6-7 sont acquises sur des fauchées de 60 km. Elles sont produites par le Centre National des Études Spatiales (CNES) et téléchargées depuis la plateforme DINAMIS (dinamis.data-terra.org/donnee).

2. Bases de données géographiques
    a) *Bases de données de l'IGN*

La BD Forêt® V2 produite par l'IGN entre 2007 et 2018 est constituée de polygones (*shapefiles*) délimitant les formations végétales de plus de 5000m² identifiées par photo-interprétation sur la France métropolitaine. Elle repose sur une nomenclature permettant de distinguer, selon un ensemble de critères, les différents types de peuplements forestiers et de végétation « basses » (ex. prairies). Cette base de données est destinée à servir de référentiel géographique forestier pour les professionnels de la filière sylvicole et acteurs de l'aménagement du territoire ou de l'environnement.

La BD ALTI® a permis d'accéder à des Modèles Numériques de Terrain (MNT) qui renseignent l'altitude du terrain. Cette donnée donne accès à la topographie qui constitue une dimension essentielle du paysage en contexte alpin.

b) Régions Bioclimatiques

L'OGM met à disposition un zonage à l'échelle des Alpes françaises des « Régions Bioclimatiques » (RBC) à l'intérieur desquelles le tétras-lyre est potentiellement présent. Ces RBC reflètent la diversité des massifs alpins à travers leurs caractéristiques climatiques, géographiques et écologiques. Les RBC peuvent être réparties en sept groupes correspondant aux différents types d'habitats potentiels du tétras-lyre, comme décrit dans le **Tableau 2** ci-dessous :

*Tableau 2. Regroupement des Régions Bioclimatiques de l'OGM en fonction des habitats privilégiés par le tétras-lyre.*

| Groupe | RBC | Habitat |
|---|---|---|
| **1** | Alpes maritimes, Alpes internes du sud | Forestier type mélézin |
| **2** | Alpes internes du Nord orientales | Forestier (essentiellement pinède) |
| **3** | Alpes internes du Nord occidentales et orientales, Alpes internes du Nord orientales – zone de transition | Landes à éricacées (rhododendrons, myrtilles, genévriers) |
| **4** | Alpes du internes du Nord occidentales – zone de transition | Habitat relativement sec, milieu de transition entre les landes à éricacées et mélézin |
| **5** | Préalpes du Nord | Pinèdes ou landes à éricacées (sol plus humide) |
| **6** | Préalpes du Sud occidentales et dépression intra-alpine du Sud, Préalpes maritimes, Plans de Provence | Habitats « atypiques » soit maigres pelouses avec pins, chênaies méditerranéennes |
| **7** | Préalpes du Sud orientales | Essentiellement prairies avec quelques pins et mélèzes, pelouses relativement sèches avec quelques pins |

### 3. Données d'observation du tétras-lyre

L'OGM a mis un place un suivi des populations du tétras-lyre à l'aide de différents protocoles :

- Observation des tétras-lyres lors de comptages au chien d'arrêt en période estivale. Ces comptages sont réalisés au sein des habitats d'élevage des nichées sur un ensemble de sites de référence et ont pour but d'estimer le succès reproducteur du tétras-lyre, à travers le comptage du nombre de « jeunes » par poule.
- Observation des tétras-lyres lors des comptages des coqs chanteurs au printemps sur un ensemble de sites de référence. Ce protocole permet notamment de produire des séries temporelles d'abondance des coqs chanteurs.
- Observation des coqs chanteurs au printemps à l'échelle des « Unités Naturelles » définies par l'OGM. L'objectif de ce protocole consiste à estimer la population du tétras-lyre à l'échelle des Alpes françaises.

### 4. Chaînes de traitement et boîtes à outils
#### a) Iota2

La chaîne de traitement *[iota2](#)*, « *Infrastructure pour l'Occupation des sols par Traitement Automatique Incorporant les Orfeo Toolbox Applications* » qui a été développée par le Centre d'Études Spatiales de la Biosphère (CESBIO) dans le but de produire des cartes d'occupation du sol (produit OSO), possède également une fonctionnalité permettant de calculer des indices phénologiques ou DHI « *Dynamic Habitat Index* » (Coops *et al.*, 2008) à partir de séries temporelles. Cette fonctionnalité repose notamment sur le travail réalisé par Hugo Trentesaux[1]. Le DHI se calcul à partir d'un indice radiométrique comme le NDVI. Tel que proposé initialement, le DHI est constitué de trois composantes :

- DHI Cum : il s'agit de la valeur cumulée de l'indice sur l'année (durée de la série temporelle). Elle peut s'assimiler à la productivité annuelle totale de la végétation.
- DHI Min : la valeur minimale de l'indice radiométrique sur l'année. Elle peut être associée au niveau minimal de productivité de la végétation.
- DHI CV : le coefficient de variation de la valeur de l'indice sur l'année (rapport entre l'écart-type et la moyenne sur l'année). Cette valeur peut s'interpréter en terme de niveau de saisonnalité.

Dans le cas de la chaîne *iota2*, des composantes alternatives sont proposées :

- **DHI Min** : ici, il s'agit du 5ème pourcentile de la série de valeurs de l'indice radiométrique considéré, sur l'année. Elle peut s'interpréter comme une valeur de productivité minimale de la végétation sur l'année.
- **DHI Max** : la valeur maximale de l'indice radiométrique sur l'année. Elle peut s'interpréter comme la valeur de productivité maximale de la végétation sur l'année.
- **DHI MOY** : la valeur moyenne de l'indice radiométrique sur l'année. Elle peut s'interpréter comme la valeur moyenne de productivité de la végétation sur l'année.

---

[1] https://gitlab.cesbio.omp.eu/trentesauxh/dhi

- **DHI STD** : l'écart-type de l'indice radiométrique sur l'année. Il peut s'interpréter comme la variation absolue de la productivité de la végétation sur l'année.
- **DHI CV** : le coefficient de variation de l'indice radiométrique sur l'année. Il peut s'interpréter comme la variation relative de la productivité de la végétation sur l'année, c'est-à-dire en termes de saisonnalité.

En plus du DHI, *iota2* permet également d'obtenir un indice correspondant au nombre de dates « claires » par pixel (autrement dit le nombre de valeurs disponibles pour calculer le DHI). Il permet de spécifier le niveau de validité des indices calculés.

La chaîne *iota2* nécessite l'installation du logiciel de gestion d'environnements python *Anaconda* ou *Miniconda*.

b) Orfeo ToolBox

La boîte à outils Open Source *Orfeo ToolBox* (OTB) fournit un large panel de méthodes permettant de réaliser une diversité de traitements sur des données d'imagerie satellitaire. OTB constitue un outil puissant et régulièrement mis à jour, pouvant être utilisé de diverses façons, notamment à travers des APIs (Python) ou dans un SIG comme le logiciel QGIS.

*5.* Package *biomod2*

Le package R [biomod2](#) (Thullier *et al.*, 2003) permet de réaliser des SDM via une approche « multimodèle ». Ce package permet de produire une modélisation centrée sur une seule espèce ou un ensemble d'espèces. L'utilisation de *biomod2* repose sur trois grandes étapes :

- **Préparation des données et paramétrage** : cette étape consiste essentiellement en un formatage des donnés et assignation des valeurs des paramètres des différents modèles. L'utilisateur doit également déterminer à cette étape quel type de pseudo-absences à prendre en compte dans la modélisation.
- **Lancement des modèles** : les modèles implémentés sont exécutés avec deux modalités possibles : séparément ou à travers un modèle d'ensemble, prenant en compte les résultats des différents modèles individuels. Différentes méthodes sont proposées pour combiner les prédictions des modèles.
- **Projection des modèles** : les prédictions de la distribution de l'espèce ou des espèces considérées sous la forme d'une probabilité de présence sont projetées sur l'emprise géographique des modèles. Cette étape permet de produire les cartes de distribution potentielles des espèces, soit en prenant en compte séparément les différents modèles, soit en considérant les prédictions combinées des différents modèles (modèle d'ensemble).

Le package *biomod2* permet d'évaluer la performance des modèles individuellement ou d'un modèle d'ensemble, à travers le calcul d'indicateurs comme la TSS (*True Skill Statistic*) ou l'aire sous la courbe ROC (*Receiver Operating Characteristic*) ou courbe de sensibilité (taux de vrais positifs) en fonction de la spécificité (taux de faux positifs). Il permet également d'évaluer l'importance relative des différentes variables.

Concernant l'utilisation des données d'observation, le package *biomod2* peut fonctionner selon deux modalités : présences *vs* « vraies absences », ou présences *vs* « pseudo-absences » (PA). Un choix important à réaliser par l'utilisateur, dans le cas où les données de « vraie absence » ne sont pas disponibles, concerne la méthode de détermination des PA et leur nombre. Ce choix dépend en partie des caractéristiques écologiques de l'espèce (ex. importance relative des contraintes géographiques *vs* gradients environnementaux) et du protocole d'acquisition des données d'observation (Barbet-Massin *et al.*, 2012).

La version 4.2-4 de *biomod2* a été utilisée pour réaliser notre modélisation.

# IV) Méthodes

## 1. Définition de la zone d'étude et des limites altitudinales

Parmi les groupes de RBC présentés dans la partie **Matériel III.2.b.**, il est admis par l'OGM que les groupes 2, 3 et 4 abritent les populations de tétras-lyre les plus importantes. Dans le cadre de notre travail, nous nous sommes concentrés sur cet ensemble comprenant : *les Alpes internes du Nord orientales, les Alpes internes du Nord occidentales, les Alpes internes du Nord orientales – zone de transition, ainsi que les Alpes internes du Nord occidentales – zone de transition* (**Figure 2**). Par soucis de commodité, nous qualifierons cet ensemble de « RBC3 ». Les habitats favorables aux tétras-lyre sont ici principalement de type forestier et des landes ligneuses à éricacées. Un filtre altitudinal a été appliqué sur le MNT, de manière à exclure de la modélisation toutes les données environnementales associées à une altitude inférieure à 1600m. En effet, d'après l'OGM, on peut considérer que les observations de tétras-lyres en dessous de 1600m dans les Alpes françaises sont devenues anecdotiques.

## 2. Calcul du Dynamic Habitat Index (DHI)

Les indices DHI ont été calculés à l'aide de la chaîne *iota2* à partir des images S2 de l'année 2020 sur l'emprise géographique de la RBC3. Nous avons sélectionné 3 indices :

- **NDVI Max** : Valeur maximale du *Normalized Difference Vegetation Index* sur l'année. Il peut s'interpréter comme le pic de croissance ou d'activité photosynthétique de la végétation.

$$\text{NDVI} = \frac{NIR - R}{NIR + R}$$

- **NDWI Max** : Valeur maximale du *Normalized Difference Water Index* sur l'année (NDWI1). Il peut s'interpréter comme le pic du contenu en eau de la végétation

$$\text{NDWI1} = \frac{NIR - SWIR}{NIR + SWIR}$$

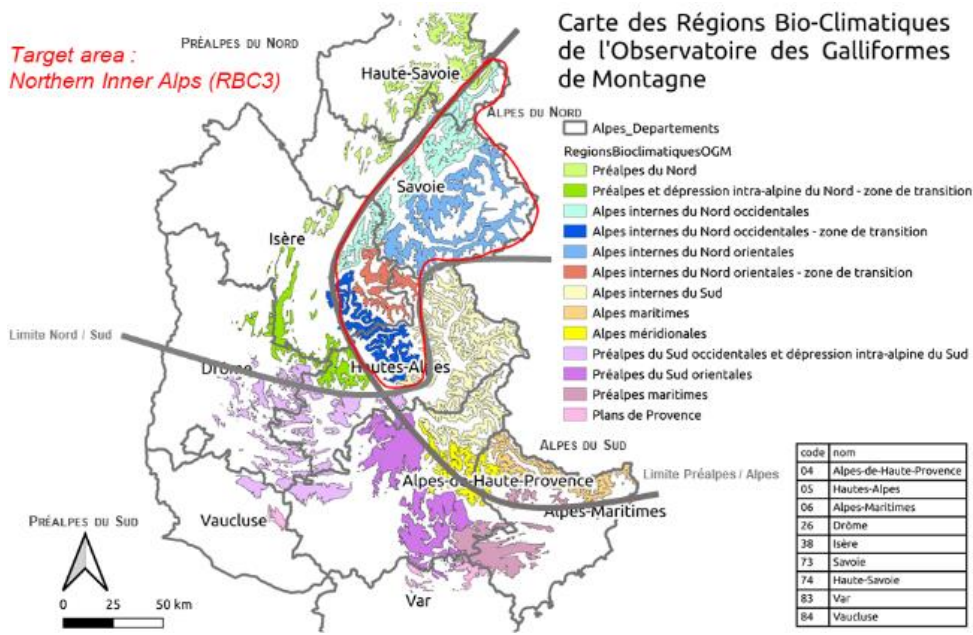

*Figure 2. Carte des Régions Bioclimatiques de l'Observatoire des Galliformes de Montagne (source : OGM). La RBC3 représentée en rouge indique l'aire géographique traitée (Alpes internes du Nord).*

Nous avons choisi de retenir uniquement la valeur maximale des indices spectraux afin d'éviter les biais potentiels induits par l'alternance de couverture neigeuse dans le calcul des autres statistiques (ex. coefficient de variation, moyenne). Le calcul du NDVI Max et du NDWI Max repose sur la fonctionnalité « *External features* » de *iota2* présente dans le [script](#) produit par Hugo Trentesaux pour le CESBIO. Elle permet de paramétrer la chaîne de traitements pour calculer une sélection d'indices spectraux dérivée de séries temporelles S2. Les formules décrivant les indices spectraux sont celles de l'OTB. Dans le cadre de notre travail, nous avons sélectionné les indices d'intérêt à partir de la mosaïque de tuiles DHI produite par lors des précédents travaux sur l'emprise de la RBC3.

### 3. Production des mosaïques à l'échelle de la RBC3

La bande panchromatique et les bandes multispectrales (Rouge, Vert, Bleu, Proche Infrarouge) des images SPOT6-7 présentes sur l'emprise de la RBC3 (année 2020) ont été utilisées afin de créer deux mosaïques d'images panchromatiques à 1,5m et « *pansharpened* » (multispectrales ramené à la résolution spectrale de l'image panchromatique par fusion). La création de ces mosaïques a nécessité une importante puissance de calcul. Elle a été réalisée sur un serveur lors des précédents travaux à l'aide des applications PYOTB (version Python des applications d'OTB).

Lors des précédents travaux, les images correspondant respectivement au DHI et au MNT ont également été assemblées pour former deux mosaïques : une mosaïque d'images DHI (NDVI Max et NDWI Max), ainsi qu'une mosaïque de MNT représentant les variations de l'altitude.

### 4. Calcul de la texture

Une série d'indices d'Haralick implémentée dans l'OTB a été calculée à partir de la mosaïque panchromatique d'images SPOT6-7. Ces indices sont calculés à partir d'une matrice de cooccurrence des niveaux de gris ou *Grey-Level Co-occurence Matrix* (GLCM). Le principe de la matrice GLCM consiste à mesurer le nombre de fois qu'un pixel d'un certain niveau de gris apparaît dans une direction définie et à une distance spécifiée depuis ses pixels voisins d'un niveau de gris spécifique (Park *et al.*, 2002). Les indices d'Haralick ont été ré-échantillonnés à 10m pour correspondre à la résolution spatiale finale des variables explicatives. Parmi les indices d'Haralick produits, nous avons retenu d'entropie qui peut s'interpréter en termes d'hétérogénéité de l'habitat. La formule de l'entropie telle que définie dans l'OTB est présentée ci-dessous :

$$\text{Entropie} = -\Sigma_{ij} g(i,j) \log_2 g(i,j) \text{ ou } 0 \; g(i,j) = 0$$

Avec $g(i,j)$ correspondant à la fréquence de l'élément indexé *i,j* dans la matrice GLCIL *(Grey Level Co-occurrence Indexed List*, une alternative à la matrice GLCM, utilisée pour augmenter la vitesse des calculs dans l'OTB)

### 5. Classification des strates de végétation

Une classification en apprentissage profond ou « *deep learning* » a été réalisée à partir de la mosaïque d'images SPOT6-7 « *pansharpened* » afin de produire une carte d'occupation du sol comprenant trois strates de végétation : « Prairies » (permanentes et temporaires), « Landes ligneuses à éricacées » et « Forêts ». En milieu alpin, ces trois strates correspondent approximativement aux strates herbacée, arbustive et arborée. Par la suite, deux classes de ligneux hauts ont été échantillonnés afin de distinguer les conifères et les feuillus. Le jeu de données d'entraînement utilisé a été constitué par échantillonnage manuel d'environ 150 polygones par classe sur les images SPOT6-7, d'une superficie variable de l'ordre de quelques hectares. La BD Forêt® V2 a été utilisée comme référence afin de faciliter l'identification des strates de végétation en prédéfinissant les zones d'échantillonnage potentiels des polygones d'entraînement. En plus des strates de végétation, trois autres classes d'occupation ont été échantillonnées : « Surfaces en eau », « Surfaces minérales » et « Surfaces ombragées » (ex. une forêt dans l'ombre d'une falaise).

La méthode de classification utilisée se décompose en deux réseaux de neurones convolutifs (CNN) appliqués successivement (**Figure 3**). Le premier CNN permet dans un premier temps de classer le pixel central à l'intérieur de chaque polygone, avant de classer chaque pixel à l'intérieur de « *patchs* » englobant les polygones d'entraînement. La classification des pixels environnants les polygones d'entraînement à l'intérieur de chaque patch apporte une information contextuelle au second CNN (« densification » de l'information) dans le but d'améliorer la gestion des zones de transitions entre les classes : on évite ainsi les effets de « lissage » des limites des différentes classes d'occupation. La classification finale d'où résulte la cartographie des strates de végétation et des autres classes d'occupation du sol est réalisée par le second CNN via une segmentation sémantique. Deux classifications ont été produites : une première avec une strate générique « Forêts » et une deuxième distinguant les types « Conifères » et « Feuillus ». La mise en place et l'application des CNN a été réalisée en étroite collaboration avec Dino Ienco (INRAE).

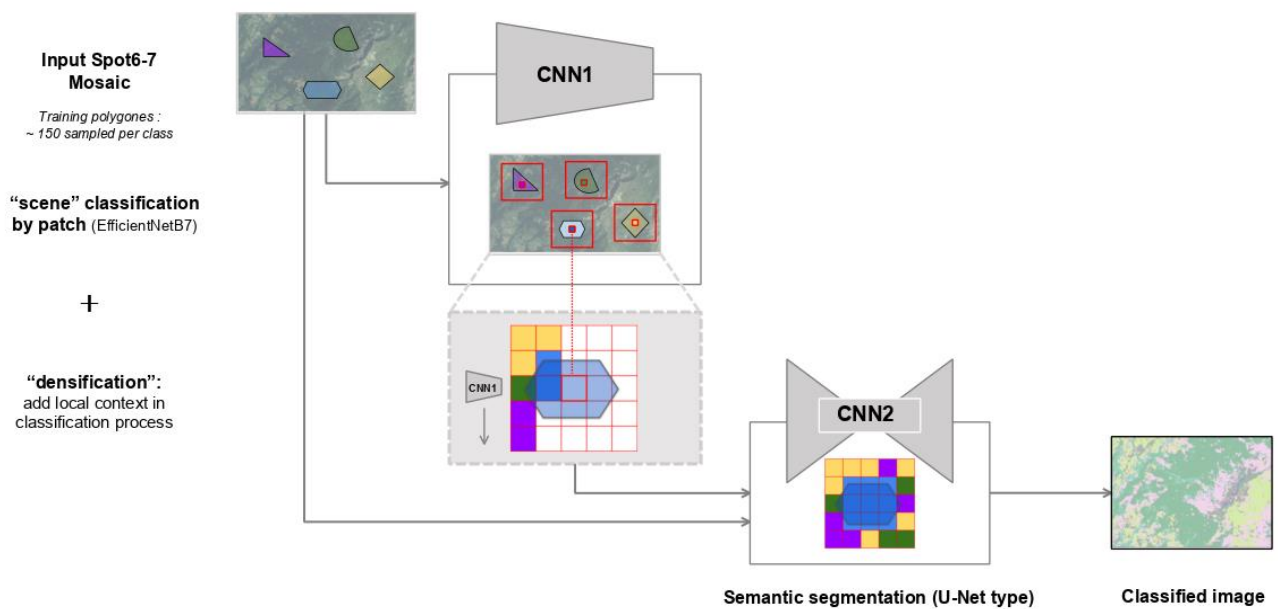

*Figure 3. Schéma simplifié du processus de classification (deep learning) de la mosaïque « pansharpened » à l'aide de deux réseaux de neurones convolutifs (CNN).*

6. Calcul du pourcentage des classes d'occupation du sol

A l'aide du logiciel R, le pourcentage de chaque classe d'occupation du sol a été déterminé à l'intérieur de cellules de 10 x 10 m correspondant aux pixels de S2. Nous faisons ici l'hypothèse que la proportion relative des différentes classes est plus informative que la simple présence-absence pour décrire la composition du paysage. Nous obtenons pour chaque classe d'occupation du sol une image avec une résolution de 100m² correspondant au pourcentage de la classe considérée. Chaque image de pourcentage de classe a été considérée comme une variable environnementale à part entière.

7. Modélisation de la distribution des nichées
    a) *Sélection des observations*

Parmi les différents jeux de données fournis par l'OGM, nous nous sommes concentrés sur le traitement des données de comptage au chien d'arrêt. Dans ce suivi, nous avons uniquement retenu les observations de nichées. Seules les observations de la période 2015-2021 ont été conservées. Les données datant d'avant 2015 ont été considérées comme potentiellement trop anciennes au regard de l'année d'acquisition des images utilisées dans notre modélisation (des changements environnementaux ont pu intervenir et avoir un impact sur la distribution des nichées).

    b) *Formatage des données et pseudo-absences*

Le cadre méthodologique du package *biomod2* nécessite un formatage particulier des données en entrée des modèles (observations et variables environnementales). Les variables environnementales ont été alignées sur une même grille correspondant à l'emprise de la RBC3 avec une taille de cellule de 10 x 10 m puis « empilées » (« stackées ») dans une image multibande. Elle constitue l'une des deux entrées de la modélisation proposée dans le cadre de *biomod2* (variables explicatives). La seconde entrée correspond à la distribution spatiale des nichées (variable à prédire). Les variables environnementales sont décrites dans le **Tableau 3.**

*Tableau 3. Liste des variables environnementales utilisées dans le SDM.*

| Nom de la variable | Description |
|---|---|
| *NDVIMax* | Indice phénologique correspondant à la valeur maximale du NDVI de l'année calculée par *iota2*. Il peut être considéré comme un proxy du pic de croissance annuel de la végétation ou du maximum de l'activité photosynthétique. |
| *NDWIMax* | Indice phénologique correspondant à la valeur maximale du NDWI1 de l'année calculée par *iota2*. Il peut être considéré comme un proxy du pic annuel du contenu en eau de la végétation. |
| *Mineral_Rate* | Proportion de surface minérale dans une cellule de 10 x 10 m obtenu par la classification en *deep learning*. |
| *Grassland_Rate* | Proportion de surface de prairies dans une cellule de 10 x 10 m obtenu par la classification en *deep learning*. |
| *Heath_Rate* | Proportion de surface de landes ligneuses à éricacées dans une cellule de 10 x 10 m obtenu par la classification en *deep learning*. |
| *Forest_Rate* | Proportion de surface de ligneux hauts dans une cellule de 10 x 10 m obtenu par la classification en *deep learning*. |
| *Haralick_Entropy* | Valeur de l'hétérogénéité spatiale ou d'entropie telle que définie par Haralick et calculé par l'OTB. |
| *Altitude* | Altitude (supérieure à 1600m) extraite du MNT fourni par l'IGN. |

Les classes de « Surfaces ombragées » et « Surfaces en eau » ont été estimées *a priori* non pertinentes pour expliquer la distribution des nichées et n'ont pas été intégrées à la liste finale des variables environnementales. Concernant la strate arbustive, la variable générique « *Forest_Rate* » a été retenue pour la réalisation du SDM. Les variables « *Deciduous_Rate* » et « *Coniferous_Rate* » dérivées de la deuxième classification n'ont pas été utilisées dans le SDM (**Annexe 1**). Dans cette seconde version de la classification, des confusions relativement importantes entre les classes « Feuillus », « Lande ligneuses à éricacées » et « Prairies » ont été observées de façon empirique. De plus, lors d'une phase de tests du SDM, les variables « *Deciduous_Rate* » et « *Coniferous_Rate* » se sont révélées peu explicatives concernant la distribution des nichées (**Annexe 2**). Pour ces différentes raisons, il est préférable de conserver la variable « *Forest_Rate* » pour la cartographie finale de la probabilité de présence des nichées (ce qui n'enlève pas le potentiel d'utilisation de cette seconde classification en tant que telle sur le plan opérationnel).

Les protocoles de suivi sur le terrain ne permettant pas de déduire de « vraies absences » des nichées, il a été nécessaire de recourir à des « pseudo-absences » (PA). Dans le cas présent, la distribution aléatoire des PA est à privilégier car nous n'avons pas émis d'hypothèses sur les différentes variables environnementales (Barbet-Massin *et al.*, 2012). Nous avons choisi d'utiliser une distribution aléatoire de 4359 PA, soit trois fois le nombre d'observations de nichées (1453) avec dix jeux de PA différents, c'est-à-dire dix répétitions (voir [les recommandations de l'équipe de développement de biomod2](#)).

Afin de produire le jeu de pseudo-absences, nous avons commencé par générer aléatoirement 43590 points aléatoires à l'intérieur de la RBC3, soit dix fois le nombre de PA souhaité. Nous avons ensuite tiré aléatoirement 4359 PA au sein de ce pool initial. Le tirage a été répété dix fois pour obtenir dix jeux de 4359 PA. Une table contenant les différents jeux de PA a été créée et utilisée dans la fonction BIOMOD_Formating() avec une définition « manuelle » des jeux de PA (*PA.strategy = 'user.defined'*, *PA.user.table = myPAtable*).

Les données en entrée de la modélisation (variables environnementales, observations, PA) ont été formatées pour être intégrées dans le cadre méthodologique de *biomod2* avec la fonction BIOMOD_FormatingData(). Dans le cas où deux observations se situaient sur la même cellule, une seule des deux a été retenue. Les PA situées sous la limite altitudinale fixée à 1600m ont été automatiquement supprimées (le nombre réel de PA utilisé dans la modélisation est donc inférieur à 4359).

*c) Présélection des variables explicatives*

Afin de tester l'existence de corrélations entre les variables environnementales, nous avons calculé le coefficient de corrélation de Pearson entre les différentes paires de variables pour un échantillon de 1000 valeurs tirées aléatoirement. La matrice de distance obtenue est utilisée pour produire une classification ascendante hiérarchique (les groupes sont formés à l'aide de la méthode de lien complet, « *complete linkage* »). Nous avons fixé une valeur seuil (*cutoff*) de regroupement des variables à 0,7 : les paires de variables présentant un coefficient de corrélation supérieur ou égal à 0,7 ont été regroupées au sein d'un même groupe. Cette étape a été réalisée à l'aide de la fonction removeCollinearity() du package R *virtual species*.

*d) Paramétrage et application des modèles*

A l'aide de la fonction BIOMOD_Modeling(), nous avons choisi de tester la totalité des 12 algorithmes disponibles dans *biomod2* (GLM, GBM, GAM, CTA, ANN, SRE, FDA, MARS, RF, MAXENT, MAXNET, XGBOOST). Les différents jeux de paramètres ont été laissés « par défaut » via la fonction BIOMOD_ModelingOptions(). Nous avons également choisi d'appliquer la méthode de validation croisée la plus simple (CV.strategy = 'random') consistant à séparer aléatoirement les données en jeux de calibration et de validation avec respectivement 80 % et 20 % des données d'origine (CV.perc = 0.8). La méthode de validation croisée a été répétée deux fois afin de diminuer la variabilité liée à la répartition aléatoire en jeux de données de calibration et de validation (CV.nb.rep = 2). Pour chaque algorithme, nous avons produit un total de 2 x 10 modèles individuels correspondant aux deux tirages utilisés pour la validation croisée et aux dix tirages aléatoires de PA.

Afin d'évaluer la performance des modèles et l'importance relative des variables environnementales, nous avons choisi d'utiliser deux métriques : la « *True Skill Statistic* » (TSS) et l'aire sous la courbe ROC *Recever Operating Characteristic* (metric.eval = c('TSS', 'ROC')). Nous appellerons cette dernière métrique « ROC » par simplicité, comme utilisé dans la terminologie de *biomod2* (pour être précis, nous devrions parler d'AUC ou *Area Under the ROC Curve*).

*e) Sélection des modèles en fonction de leur performance relative*

La performance des algorithmes a été évaluée par les métriques de la ROC et du TSS à l'aide des fonctions get_evaluations() et bm_PlotEvalBoxplot(). Cette étape a permis de sélectionner le ou les algorithmes pertinents pour prédire la distribution des nichées. A l'aide des fonctions get_variables_importance() et bm_PlotVarImpBoxplot() l'importance relative des variables environnementales a également été examinée.

*f) Réalisation du modèle d'ensemble et projection des prédictions*

Après examen de la performance des algorithmes, le Random Forest (RF) a été sélectionné (voir détails partie **Résultats V.5.b.**). Un jeu de PA plus restreint a été produit afin de limiter la puissance de calcul nécessaire à la modélisation tout en suivant les recommandations de l'équipe de développement de *biomod2*, soit autant de PA que d'observations (recommandé dans le cas des algorithmes CTA, BRT et RF). A partir du pool initial de 43590 PA, nous avons tiré trois jeux aléatoires de 1453 PA soit autant de PA que d'observations. Au total, 2 x 3 modèles individuels basés sur le RF ont été produits (nombre de tirages utilisés pour la validation croisée par le nombre de répétitions de PA) avec la fonction BIOMOD_Modeling(). Comme précédemment, les PA situés en dessous de 1600m d'altitude ont été filtrés. De même, pour deux observations situées sur une même cellule, une seule aura été retenue.

Les six modèles RF ont été combinés pour former un modèle d'ensemble à l'aide la fonction BIOMOD_EnsembleModeling(). Le modèle d'ensemble est utilisé afin de prendre en compte les résultats des modèles individuels et réduire la variabilité associée à chacun d'eux (ici la variabilité due aux tirages de PA et au partitionnement de la validation croisée). La médiane a été utilisée pour combiner les prédiction des modèles individuels de RF (em.algo = c('EMmedian')).

Les modèles individuels et le modèle d'ensemble ont été projetés sur l'emprise de la RBC3 au-dessus de 1600m d'altitude à l'aide des fonctions BIOMOD_Projection() et BIOMOD_EnsembleForecasting(). La cartographie de la probabilité de présence des nichées du tétras-lyre a été obtenue à l'échelle de la RBC3. Les valeurs de probabilités sont comprises dans une échelle de 0 à 1000 pour des raisons de gestion de la RAM.

# V) Résultats

## 1. Mosaïques panchromatique et multispectrale

Les mosaïques d'images SPOT6-7 panchromatiques et « *pansharpened* » (multispectrales) qui ont été obtenues permettent de couvrir la majorité des Alpes françaises (notamment la RBC3) avec une résolution de 1,5 m (**Figure 4**). On notera que les zones suivantes n'ont pas été intégrées aux mosaïques : les extrémités sud et nord des pré-Alpes du nord, la partie ouest des pré-Alpes du Sud et l'extrémité sud-est des Alpes maritimes (non nécessaires pour le traitement de la RBC3).

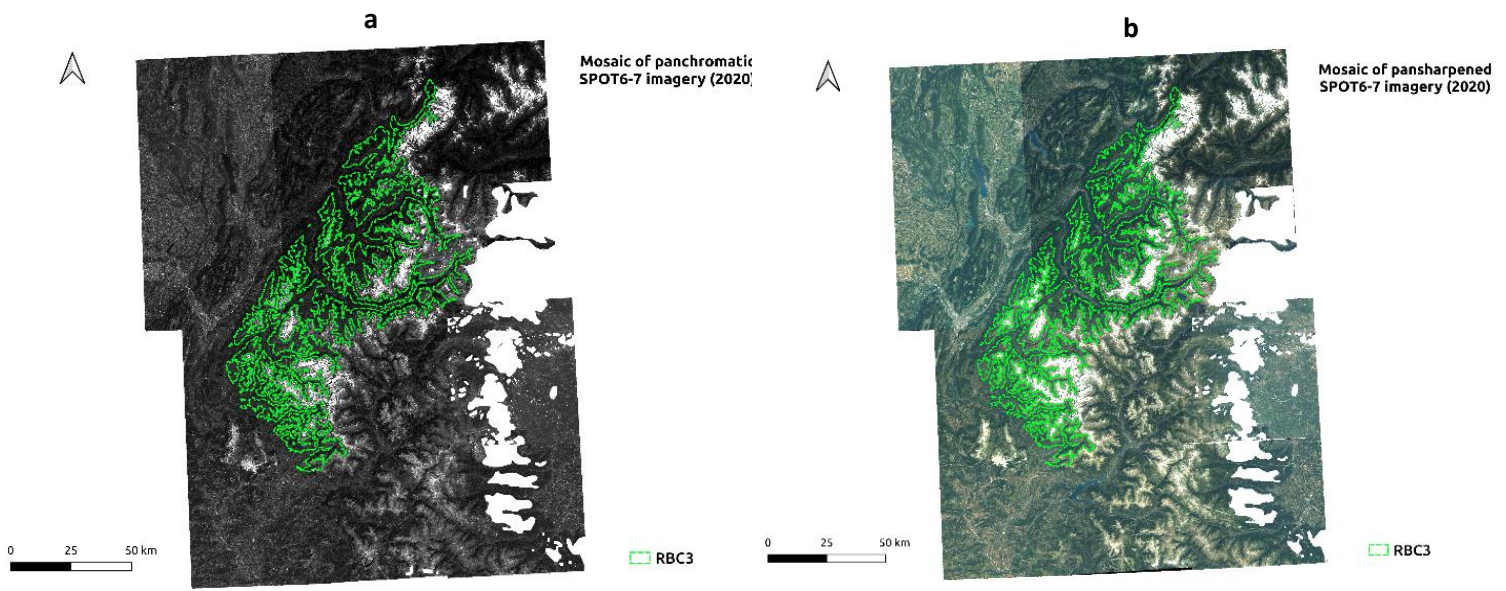

***Figure 4.*** *Mosaïques d'images SPOT6-7 à l'échelle des Alpes françaises (2020).* ***a)*** *Images panchromatiques (niveaux de gris).* ***b)*** *Images « pansharpened » (multispectrales), rééchantillonnées à 1,5m soit la résolution des images panchromatiques. Les pointillés verts indiquent les limites de la RBC3.*

## 2. Variations du Dynamic Habitat Index

La **Figure 5** montre un exemple de la distribution spatiale du NDVI Max (DHI) autour du lac de la Girotte (Savoie). Dans la suite du rapport, nous continuerons de donner cette zone en exemple. Il s'agit de l'une des zones d'étude utilisée dans le cadre des précédents travaux. De plus, des nichées de tétras-lyre ont été observées à l'ouest du lac, au sein d'un environnement qui peut facilement être interprété sur les images en RVB (différenciation entre les landes ligneuses à éricacées, forêts et prairies). La distribution du DHI à l'échelle de la RBC3 peut être visualisée en **Annexe 3**.

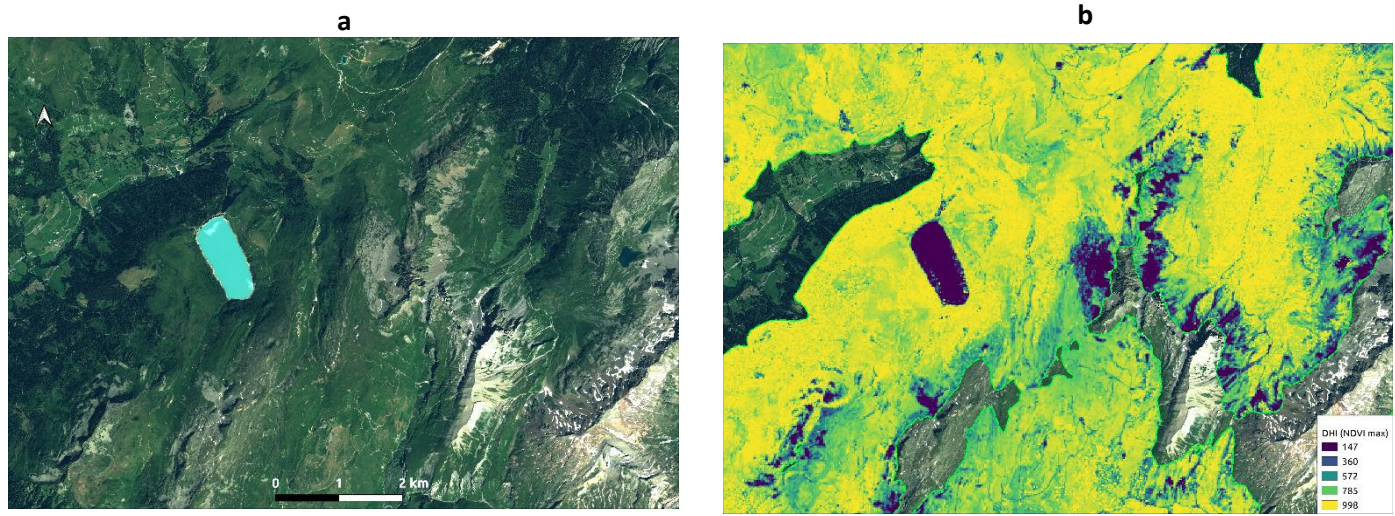

*Figure 5. Distribution du pic de productivité annuel par pixel à proximité du lac de la Girotte (2020). **a)** Mosaïque RVB des images SPOT6-7 « pansharpened » (multispectrale). **b)** Variations spatiales du NDVI Max (DHI). Les pointillés indiquent les limites de la RBC3.*

3. Hétérogénéité du paysage : entropie d'Haralick

Après un examen visuel, nous avons pu observer que l'augmentation des valeurs de l'entropie d'Haralick dans la RBC3 est essentiellement liée à la présence de formations rocheuses, d'éboulis et de structures d'origine anthropique (routes, bâtiments…). Un exemple de ces variations est visible sur la **Figure 6**.

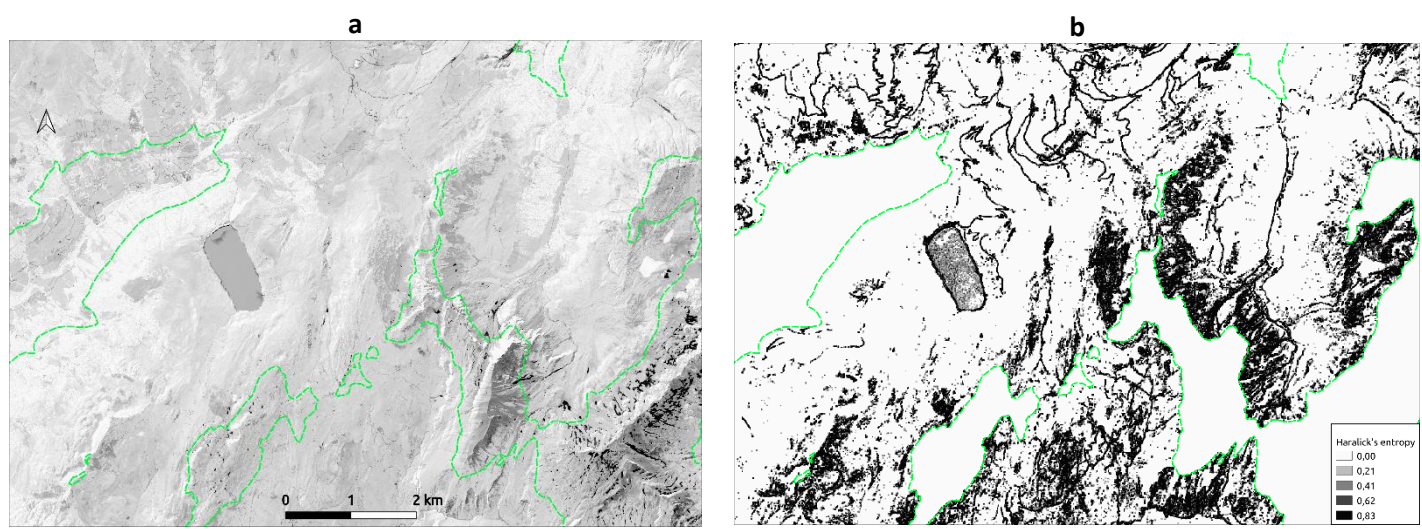

*Figure 6. Hétérogénéité du paysage à proximité du lac de la Girotte (2020). **a)** Mosaïque des images SPOT6-7 panchromatiques. **b)** Entropie d'Haralick calculée à l'aide de l'OTB. Les pointillés verts indiquent les limites de la RBC3.*

4. Strates de végétation

Après examen du produit de la classification (**Figure 7**), nous avons pu observer une différenciation relativement précise des classes avec des confusions réduites (F-Score ~ 85 %). Les confusions les plus importantes identifiables graphiquement sont entre les classes « *Heaths* » (landes ligneuses à éricacées) et « *Grassland* » (prairie).

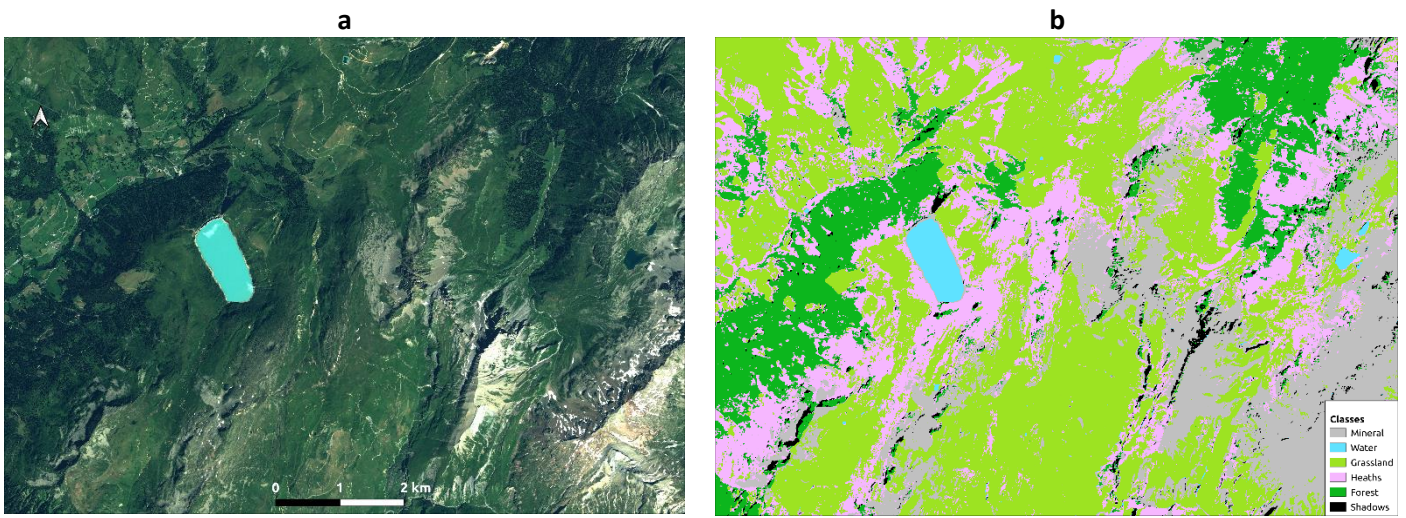

***Figure 7.*** *Classification de l'occupation du sol obtenue à proximité du lac de la Girotte (2020). **a)** Mosaïque RVB des images SPOT6-7 « pansharpened » (mutlispectrale). **b)** Produit de la classification issue des deux CNN successifs (« deep learning »).*

5. Distribution des nichées

a) Examen des corrélations entre les variables environnementales

L'examen de la classification ascendante hiérarchique (CAH) des variables environnementales (**Figure 8**) révèle une absence de corrélation suffisamment importante pour produire des groupes de variables avec un seuil défini à 0,7 (*Pearson's r*). Nous avons donc choisi de retenir toutes les variables environnementales pour prédire la probabilité de présence des nichées.

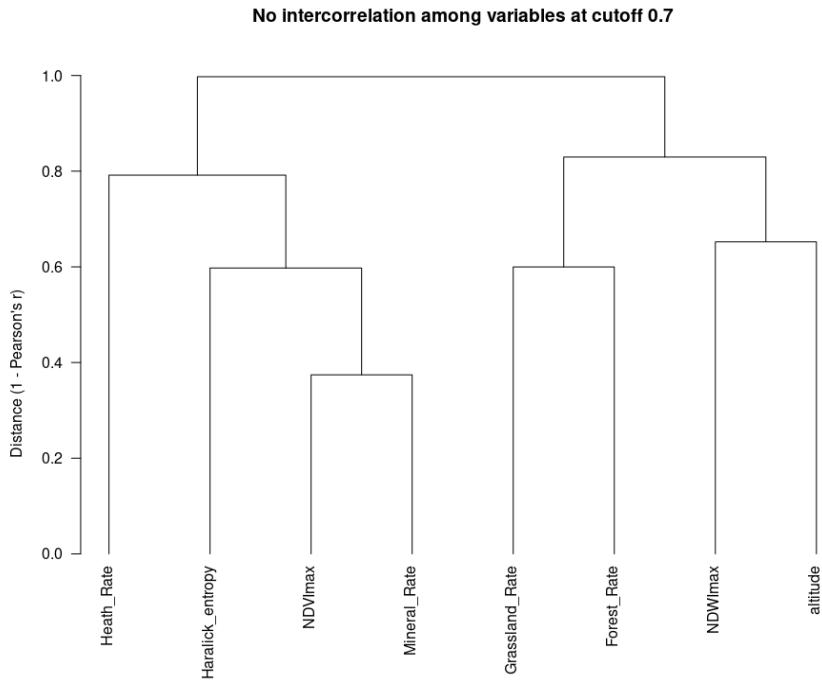

*Figure 8. Intercolinéarité des variables environnementales. Arbre de classification ascendante hiérarchique (CAH) représentant l'inter-corrélation des variables environnementales utilisées dans le SDM.*

b) *Performance des modèles*

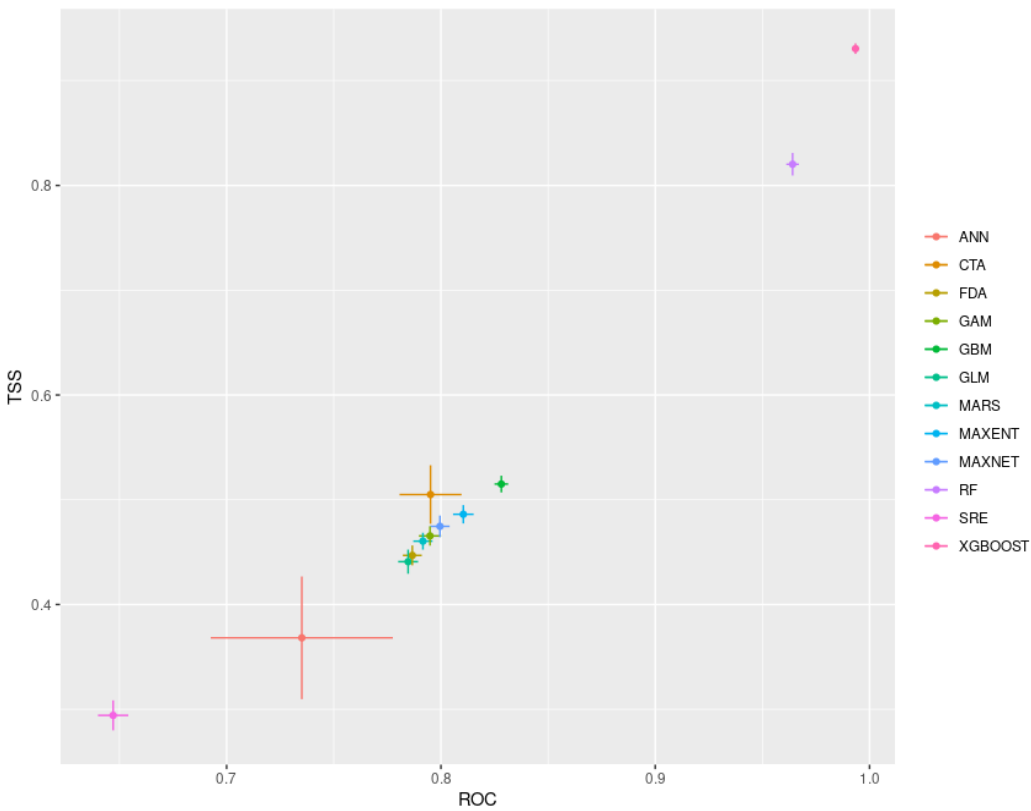

*Figure 9. Distribution de la performance des algorithmes utilisés dans les modèles individuels. La performance est estimée à l'aide de l'aire sous la courbe ROC (Receiver Operating Charateristic) en abscisse et de la True Skill Statistic (TSS) en ordonnée.*

Sur la **Figure 9** nous pouvons observer que deux algorithmes possèdent une performance très élevée d'après les métriques ROC et TSS : le *Random Forest* (RF) (ROC = 0,96 ; TSS = 0,82) et l'*Extreme gradient boosting* (XGBOOST) (ROC = 0,99 ; TSS = 0,93). Les autres algorithmes (CTA, FDA, GAM, GBM,

GLM, MARS, MAXENT, MAXNET, SRE, ANN) montrent des performances nettement plus faibles (ROC < 0,83 ; TSS < 0,52). Nous avons choisi de retenir uniquement le RF pour la réalisation du SDM en raison de sa sensibilité théoriquement plus faible aux effets de sur-apprentissage et son paramétrage plus simple que le XGBOOST.

### c) *Importance relative des variables environnementales*

Nous pouvons observer que l'altitude, le taux de landes ligneuses à éricacées et le DHI (NDVI Max et NDWI Max) constituent les principales variables explicatives des modèles basés sur le RF (**Figure 10**). Dans le cas des modèles individuels (**Figure 10-a**), l'altitude a une importance de ~25 %, le taux de landes ligneuses à éricacées 23 % (avec un écart-type supérieur) et le DHI entre 10 et 15 %. Dans le cas du modèle d'ensemble (**Figure 10-b**), l'altitude a une importance de 24 %, le taux de landes ligneuses à éricacées de ~14 % et le DHI autour de 15 %. Les écart-types associés aux variables environnementales (à l'exception du taux de landes ligneuses à éricacées) sont relativement réduits. Ils montrent une influence limitée des jeux de PA et de partitionnement des données lors de la validation croisée sur les prédictions des nichées.

En **Annexe 4** sont présentés les résultats de l'importance relative des variables environnementales pour les différents algorithmes disponibles dans *biomod2*.

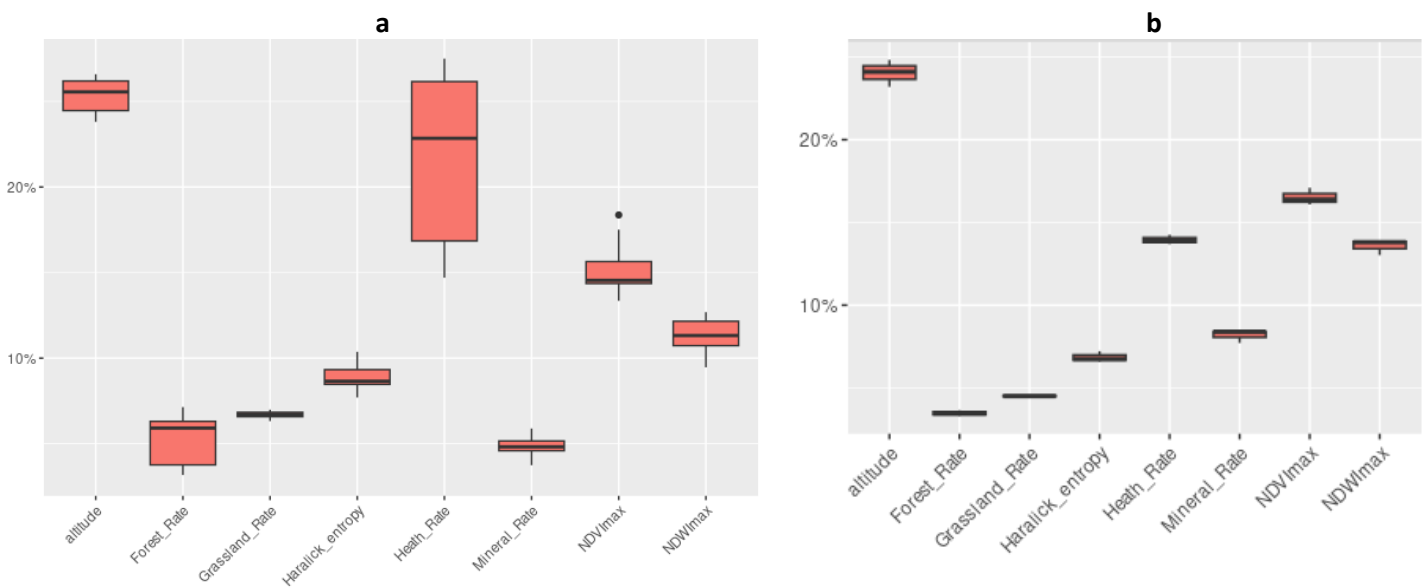

*Figure 10. Importance relative des variables environnementales des modèles basés sur le Random Forest. a) Variabilité associée aux six modèles individuels. b) Variabilité associée au modèle d'ensemble (médiane des prédictions des modèles individuels). L'importance relative est obtenue par le calcul suivant :*
1 – PearsonCor(OriginalVariable, ShuffledVariable), voir *https://rdrr.io/rforge/biomod2/man/variables_importance.html*.

d)   *Cartographie de la probabilité de présence des nichées*

La **Figure 11** montre les variations de la probabilité de présence des nichées du tétras-lyre autour du lac de la Girotte à partir du modèle d'ensemble. Nous pouvons observer une influence de l'altitude à travers une diminution nette de la probabilité de présence dans la partie supérieure de la RBC3 (entre 2300 et 2400 m d'altitude environ après examen du MNT) avec une probabilité proche de zéro (**Figure 11 – a – c**). D'autre part, nous pouvons également observer l'influence positive des landes ligneuses à éricacées sur la probabilité de présence des nichées avec des valeurs > 900 (sur une échelle de 0 à 1000) (**Figure 11 – b – d**).

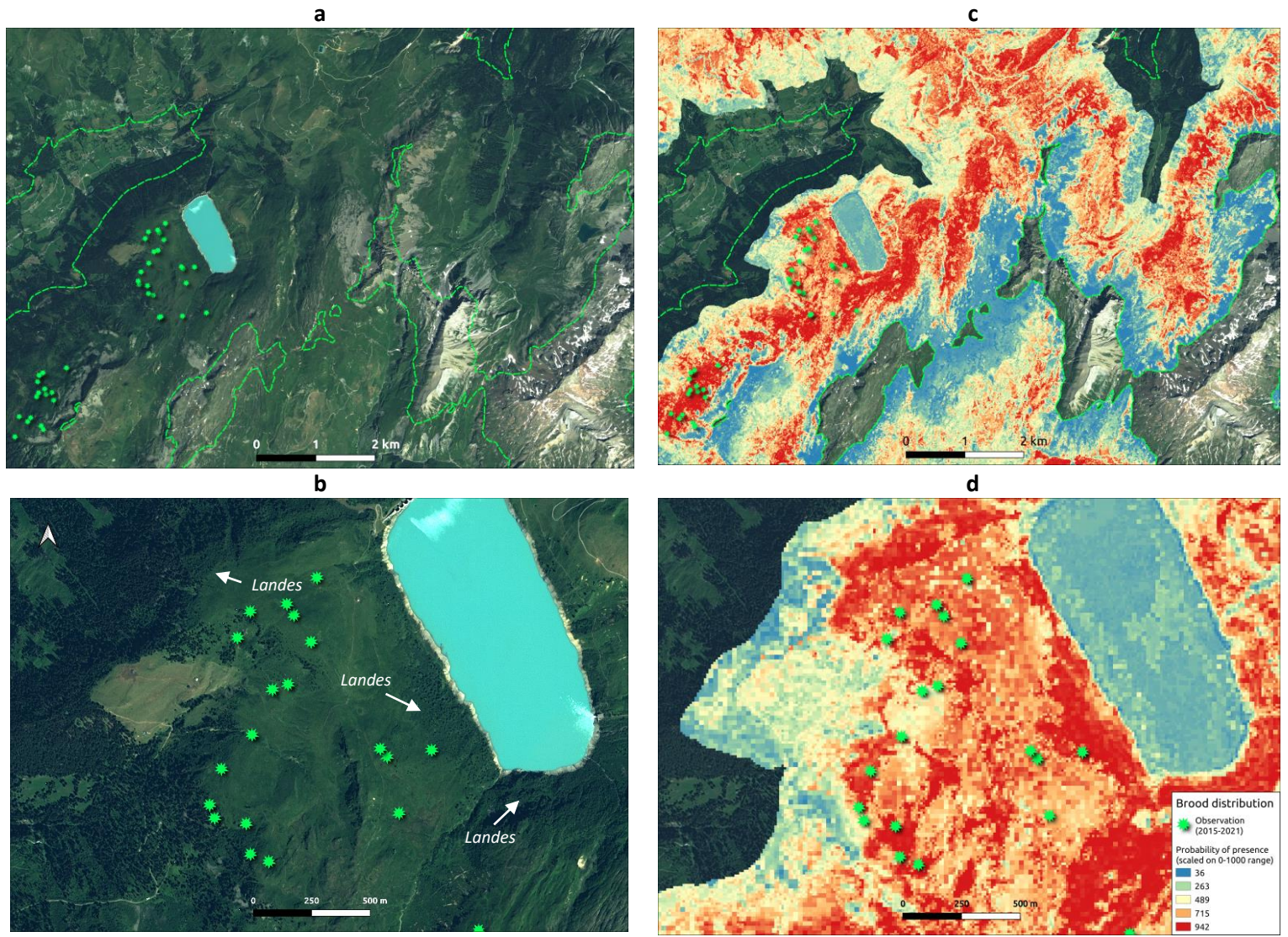

***Figure 11.*** *Projection de la probabilité de présence des nichées du tétras-lyre à proximité du lac de la Girotte (2020). **a) b)** Cartes RVB d'images SPOT6-7 multispectrale (« pansharpened »). **c) d)** Probabilité de présence des nichées exprimée sur une échelle 0-1000. Cette probabilité a été déterminée à partir du modèle d'ensemble (médiane) de 20 modèles individuels de RF correspondant aux deux tirages utilisés pour la validation croisée et aux 10 tirages aléatoires de PA. Les pointillés verts représentent les limites de la RBC3 entre 1400 et 2400 m d'altitude.*

# VI) Discussion

## 1. Performance et paramétrage des modèles

Nous avons observé que les algorithmes d'apprentissage automatique ou « *machine learning* » RF et XGBOOST présentent une performance nettement supérieure à celle des autres algorithmes, y compris l'ANN (apprentissage profond ou « *deep learning* ») (**Figure 9**). Toutefois, dans le cadre de notre travail, nous avons utilisé les jeux de paramètres « par défaut » de *biomod2*. Un travail sur le paramétrage des différents algorithmes (hyperparamètres dans le cas de l'ANN) devrait permettre d'améliorer leurs performances respectives. Dans le cas du XGBOOST, les scores de TSS et ROC proches de 1 (0.93 et 0.99 respectivement) pourraient s'expliquer par un effet de sur-apprentissage. Bien que le RF soit en théorie relativement moins sensible au sur-apprentissage que le XGBOOST, il pourrait être intéressant de tester d'autres méthodes de validation croisée afin de vérifier l'effet de la méthode sur la performance des modèles, ainsi que leur « transférabilité géographique » comme le propose les méthodes « *block* » (Muscarella et al. 2014) et « *stratified* » (Wenger and Olden 2012) de *biomod2*. De premiers tests ont été réalisés avec ces deux méthodes (non présentés dans ce rapport). Les performances du RF et du XGBOOST n'ont été que légèrement impactées. Le résultat de ces premiers tests tend à montrer une absence d'effet de sur-apprentissage et de dépendance vis-à-vis du contexte géographique local. Toutefois, une analyse plus profonde des effets du paramétrage et de l'impact du choix de la méthode de validation croisée devrait être menée dans le but de proposer une modélisation la plus rigoureuse possible.

Dans le cadre de notre travail, nous avons cherché à nous rapprocher autant que possible d'une sélection aléatoire de PA. Le tirage aléatoire de PA dans le pool initial de PA tel qu'il a été réalisé n'exclut pas en théorie de réutiliser plusieurs fois certaines PA (équivalent d'un tirage avec remise). Étant donné que le nombre de PA dans le pool initial (43590) est 30 fois supérieur au nombre de PA par tirage, la probabilité de tirer deux fois une même PA demeure néanmoins très faible (0,09% si les 1453 PA sont utilisées). La méthode appliquée ici présente une alternative théoriquement plus simple pour l'utilisateur : définir directement un nombre de PA aléatoires à l'aide de la fonction BIOMOD_Formating(). Dans notre cas, il n'a pas été possible d'utiliser cette fonctionnalité (PA.strategy = 'random') en raison d'un problème de saturation de la mémoire de R. Cette difficulté, qui n'a pu être entièrement contournée (des tests ont été réalisés sur un serveur) indique que *biomod2* peut nécessiter une importante puissance de calcul, voir la mise en place d'une parallélisation des traitements. De futurs travaux proposant une modélisation à grande échelle de la distribution des nichées doit prendre en compte cette problématique de gestion de la puissance de calcul et de la RAM à mobiliser.

## 2. Influence des variables environnementales

Les résultats du SDM confirme l'importance des landes ligneuses à éricacées pour cartographier l'habitat de reproduction du tétras-lyre. L'importance du NDVI Max (corrélé à la productivité primaire et la verdeur de la végétation) et du NDWI Max (corrélé au contenu en eau de la végétation) pour expliquer la probabilité de présence des nichées pourrait s'expliquer par une relation entre ces indices et l'existence de « pics » locaux en termes d'abondance ou de qualité des ressources alimentaires. Toutefois, l'altitude demeure la variable la plus importante du modèle d'ensemble pour prédire cette probabilité de présence. Nous pouvons expliquer cette influence à travers *(i)* l'existence d'une zone de transition entre la forêt (limite supérieure des arbres) et des habitats plus ouverts favorables aux tétras-lyre et *(ii)* la présence de formations rocheuses ou enneigées à partir d'une altitude proche de 2400m (limite altitudinale supérieure de la RBC3). L'altitude a donc un effet qui peut être positif ou négatif sur la probabilité de présence des nichées. De manière assez surprenante, nous n'avons pas observé de corrélation significative entre l'altitude et le taux de landes ligneuses à éricacées (**Figure 8**). Ce fait peut s'expliquer par la tendance de la classification à intégrer dans la classe « landes ligneuses à éricacées » d'autres types de végétation non échantillonnés et qui ne sont pas nécessairement favorables à la présence de nichées (ex. Aulnaies). Nous pouvons également faire l'hypothèse que les variations altitudinales de la limite supérieure des arbres dans les Alpes (entre 2200 et 2400m selon les versants nord et sud respectivement) soit suffisamment importante pour limiter cette corrélation.

Inversement, les variables issues de la classification « taux de forêt », « taux de prairie », « taux de minéral » expliquent faiblement la probabilité de présence des nichées. Dans le cadre de notre modélisation, nous avons choisi de définir une grille avec des cellules de 10 x 10 m pour calculer les taux des classes d'occupation du sol (soit la résolution finale des rasters en entrée du SDM). Il est possible que cette taille de cellule, en dehors de « fines » zones de transition, ne permette pas de mettre en évidence une influence de ces classes sur la probabilité de présence des nichées. Autrement dit, cette taille de cellule se rapproche d'une intégration des classes sous forme de présence/absence et ne permet peut-être pas de capter l'influence de la composition du paysage. Toutefois, déterminer la résolution spatiale la plus adaptée pour caractériser la composition du paysage n'est pas trivial. Il aurait été intéressant de tester différentes tailles de cellules (ex. 50, 100m). L'entropie d'Haralick explique également une part relativement plus faible de la probabilité de présence des nichées. Il est possible que l'entropie d'Haralick soit insuffisante pour représenter l'hétérogénéité du paysage ou que cette hétérogénéité soit d'une importance secondaire pour les nichées du tétras-lyre.

### 3. Validation des cartographies par regard d'expert

Deux produits cartographiques constituent des résultats potentiellement utilisables sur le plan opérationnel par l'OGM : la classification de l'occupation du sol (avec et sans distinction de la classe « Forêt » en « Feuillus » et « Résineux ») et la carte de probabilité de présence des nichées (proxy de l'habitat de reproduction du tétras-lyre) à l'échelle de la RBC3. Ces deux produits ont été soumis à l'examen de Marc Montadert, avec une attention particulière sur la station de Méribel. Il est ressorti que *(i)* les deux versions de la classification sont cohérentes avec les connaissances de terrain sur la station de Méribel *(ii)* les probabilités de présence reflètent de manière cohérente les préférences du tétras-lyre en terme d'habitat de reproduction. Les types d'erreurs ou d'imprécisions les plus évidents après examen visuel concernent l'existence de « faux positifs » évidents au sein de la classe « Feuillus » et une tendance probablement trop « optimiste » du modèle d'ensemble à prédire la présence des nichées. Le retour mentionné ici demeure qualitatif et concerne une emprise géographique limitée. Cependant, il permet d'apporter une première confirmation du potentiel d'utilisation de ces produits cartographiques sur un plan opérationnel. Concernant le mode de visualisation du SDM, il a été recommandé d'appliquer un seuillage afin de visualiser uniquement les zones avec une forte probabilité de présence des nichées, par exemple ≥ 80 %. Dès lors, il devient facile de transformer cette représentation en carte de présence/absence des nichées du tétras-lyre.

## VII) Conclusion et Perspectives

Notre travail confirme l'importance des landes ligneuses à éricacées en tant qu'abri et source de nourriture pour prédire la probabilité de présence des nichées. Dans une démarche visant à poursuivre la cartographie de l'habitat de reproduction du tétras-lyre, les variables liées à la présence de nourriture ainsi que l'altitude seront probablement les plus utiles.

Le travail que nous avons réalisé pourrait néanmoins être amélioré pour la poursuite des efforts concernant le paramétrage du SDM et la préparation des variables environnementales. Concernant le SDM, en cas de maintien d'une performance élevée du XGBOOST et d'absence d'effet important de sur-apprentissage, il serait intéressant de tester un modèle d'ensemble combinant les prédictions du RF et du XGBOOST. L'intégration des prédictions de cet algorithme pourrait contribuer à rééquilibrer l'importance relative donnée à la variable « taux de landes ligneuses à éricacées » qui rend probablement la cartographie finale trop « optimiste ». La préparation des variables environnementales pourrait également être améliorée, notamment à travers le test de différentes résolutions de grille pour caractériser la composition du paysage (taux de surface des classes d'occupation du sol). Enfin, il serait intéressant et utile pour l'OGM de produire des SDM centrés sur les autres RBC, notamment les RBC 1 et 4. En effet, selon aires géographiques, les préférences écologiques du tétras-lyre peuvent varier. C'est particulièrement le cas dans les Alpes du Sud où le mélézin revêt une importance particulière pour des raisons qui ne sont pas encore bien connues (ex. abri contre les prédateurs, recherche de fraîcheur...).

Ce travail a permis de jeter les bases d'une cartographie de l'habitat de reproduction du tétras-lyre à l'échelle des Alpes françaises. Une poursuite de l'étape de validation par les experts de l'OGM permettrait d'évaluer plus finement son utilisation potentielle sur le plan opérationnel. Conformément aux objectifs de l'OGM, la distribution spatiale de la probabilité de présence des nichées devrait pouvoir servir de variable explicative à la probabilité de présence des coqs chanteurs. Elle contribuerait ainsi à améliorer la stratification du futur plan d'échantillonnage pour le suivi des coqs chanteurs.

## VIII) Livrables

Dans le **Tableau 4** sont énumérés les livrables fournis à l'Observatoire des Galliformes de Montagne. Le principal livrable est l'image de la distribution de la probabilité de présence des nichées du tétras-lyre. Elle peut servir en tant que cartographie de son habitat de reproduction à l'échelle de la Région Bioclimatique 3 (voir **Figure 2**). Les deux produits de la classification constituent également un résultat important et peuvent être utilisés en dehors du SDM. Les mosaïques d'images SPOT6-7 panchromatiques et *pansharpenend* (multispectrales) peuvent se révéler utiles à la visualisation des paysages alpins. Enfin, les variables d'entrée du SDM sont fournies (image des variables environnementales, observations des nichées agrégées) ainsi que le script d'exécution en langage R.

*Tableau 4. Liste des livrables fournis à l'Observatoire des Galliformes de Montagne.*

| Nom | Format et taille | Description |
|---|---|---|
| CLassif_DL_b7_v4_dense_1-6 | Raster (TIF) 17,5 Go | Produit d'une classification de l'occupation du sol à l'échelle des Alpes française issue de réseaux de neurones convolutifs (*deep learning*). Six classes ont été déterminées : « Surfaces Minérales » (1), « Surfaces en eau » (2), « Prairies » (3), « Landes ligneuses à éricacées » (4), « Forêts » (5), « Surfaces ombragées » (6). *N.B. Classification utilisée dans le SDM.* |
| CLassif_DL_b7_v4_dense_1-7 | Raster (TIF) 17,5 Go | Produit d'une classification de l'occupation du sol à l'échelle des Alpes française issue de réseaux de neurones convolutifs (*deep learning*). Sept classes ont été déterminées : « Surfaces Minérales » (1), « Surfaces en eau » (2), « Prairies » (3), « Landes ligneuses à éricacées » (4), « Feuillus » (5), « Conifères » (6), « Surfaces ombragées » (7). *N.B. Classification non utilisée dans le SDM.* |
| EnvVar_Stack_RBC3_rs_min1600m | Raster (TIF) 725 Mo | Image contenant les variables environnementales utilisées dans le SDM. Elle contient les bandes suivantes : *NDVIMax, NDWIMax, Mineral_Rate, Grassland_Rate, Heath_Rate, Forest_Rate, Haralick_Entropy, Altitude.* La description des différentes bandes peut être retrouvée dans le **Tableau 3.** |
| proj_RF_RandomPA_AllModels_Lyrurustetrix_ensemble | Raster (TIF) 88 Mo | Image de la probabilité de présence des nichées du tétras-lyre à l'échelle de la RBC3 sur une échelle de 0 à 1000 obtenue par le SDM (médiane des prédictions de modèles *Random Forest*). Elle constitue, avec les produits de la classification, le résultat principal de notre travail. La probabilité de présence a été déterminée sur une tranche altitudinale comprise entre 1600 et 2400m. |
| PAN_mosaic_RBC3 | Raster (TIF) 35,1 Go | Mosaïque d'images SPOT6-7 panchromatiques à 1,5 m de résolution spatiale, de l'année 2020, à l'échelle des Alpes françaises. |
| PXS_mosaic_RBC3 | Raster (TIF) 140 Go | Mosaïque d'images SPOT6-7 *pansharpened* (multispectrales) à 1,5 m de résolution spatiale, de l'année 2020. Les bandes sont : Bleu, Vert, Rouge, Proche Infrarouge à l'échelle des Alpes françaises. |
| TETRAS_nichees_2015-2021_15000_randomPA_merged | Vecteur (shp) 1 Mo | Points d'observation des nichées du tétras-lyre par comptage au chien d'arrêt (données agrégées des années 2015-2021) et pseudo-absences. |
| Run_SDM_biomod2_RF_RBC3 | Fichier texte (.R) 16,5 Ko | Script en langage R commenté permettant d'exécuter le SDM avec le package *biomod2.* Le script est composé d'une partie permettant de préparer les variables environnementales en entrée du SDM et de les regrouper dans le format utilisé par *biomod2*. Il permet également de vérifier l'absence de corrélations entre les variables environnementales. La seconde partie du script suit la trame proposée par l'équipe de développement de *biomod2*, du formatage initial des données à la production de la carte de probabilité de présence. |

# IX) ANNEXES

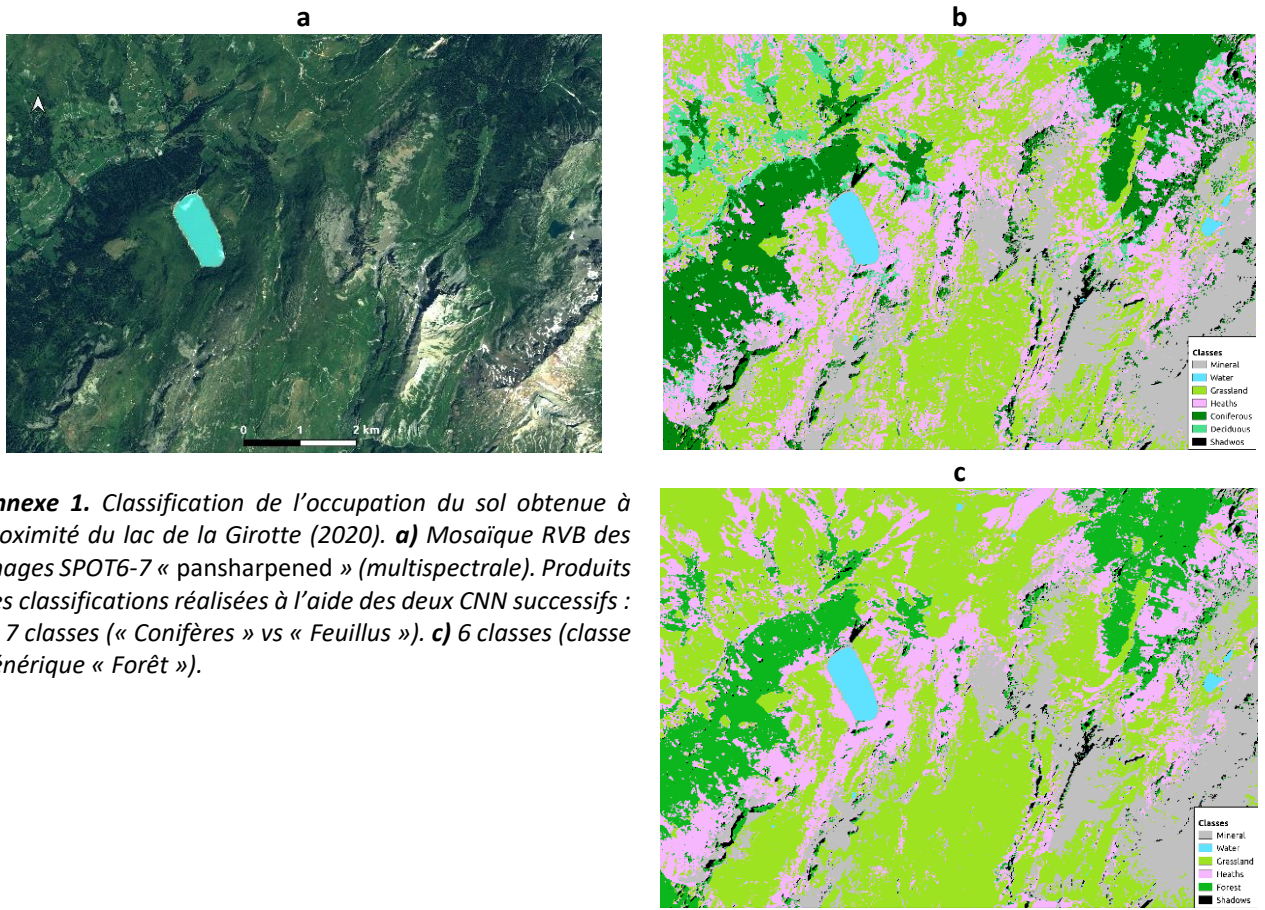

***Annexe 1.*** *Classification de l'occupation du sol obtenue à proximité du lac de la Girotte (2020). **a)** Mosaïque RVB des images SPOT6-7 « pansharpened » (multispectrale). Produits des classifications réalisées à l'aide des deux CNN successifs : **b)** 7 classes (« Conifères » vs « Feuillus »). **c)** 6 classes (classe générique « Forêt »).*

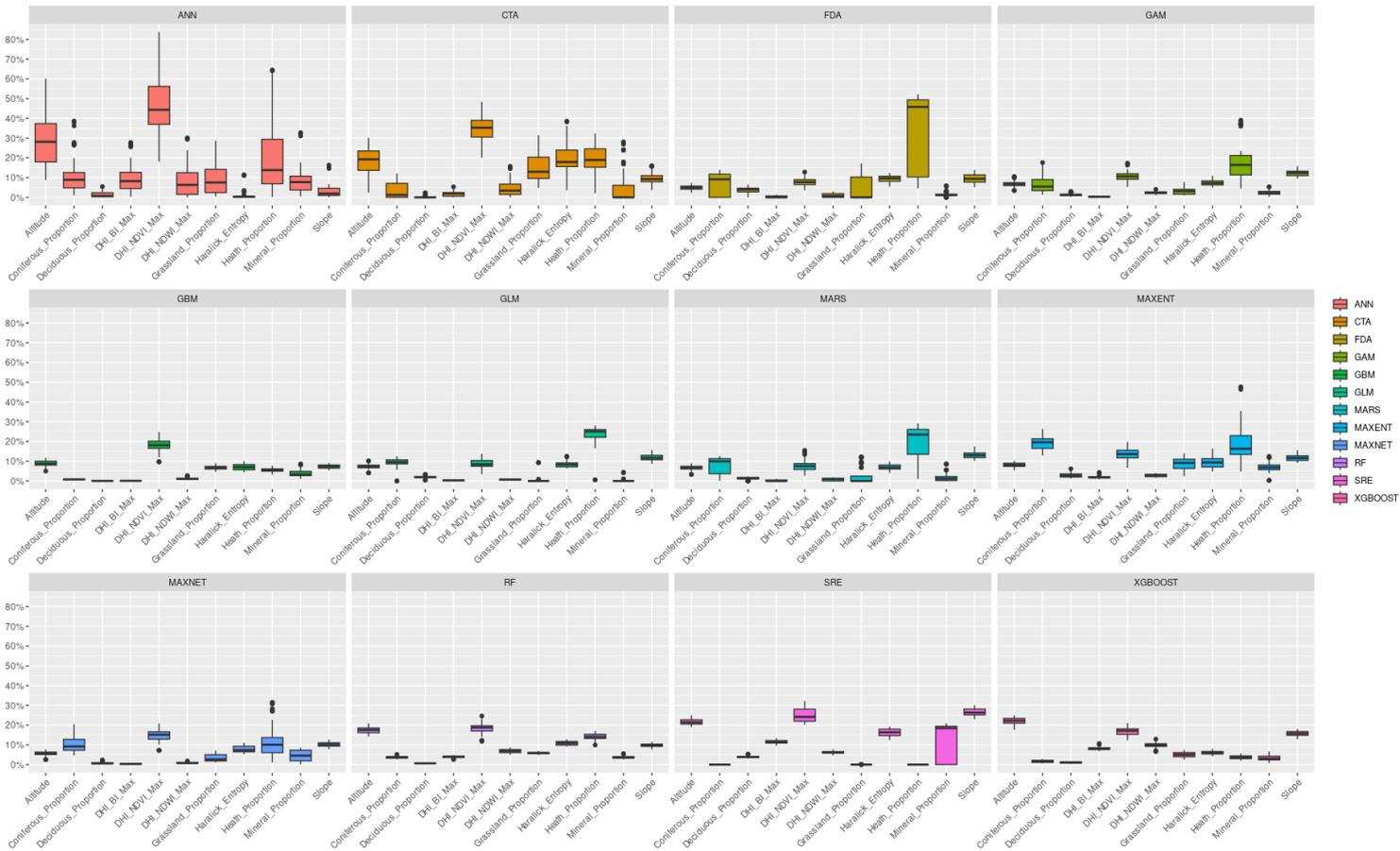

***Annexe 2.*** *Importance relative des variables environnementales du SDM distinguant les classes «* Deciduous_Rate *» et «* Coniferous_Rate *» pour chaque algorithme des modèles individuels. Dans le cas du RF on remarque une importance relative quasi nulle de la variable «* Deciduous_Rate *» (~0%) et très faible dans le cas de la variable «* Coniferous_Rate *».*

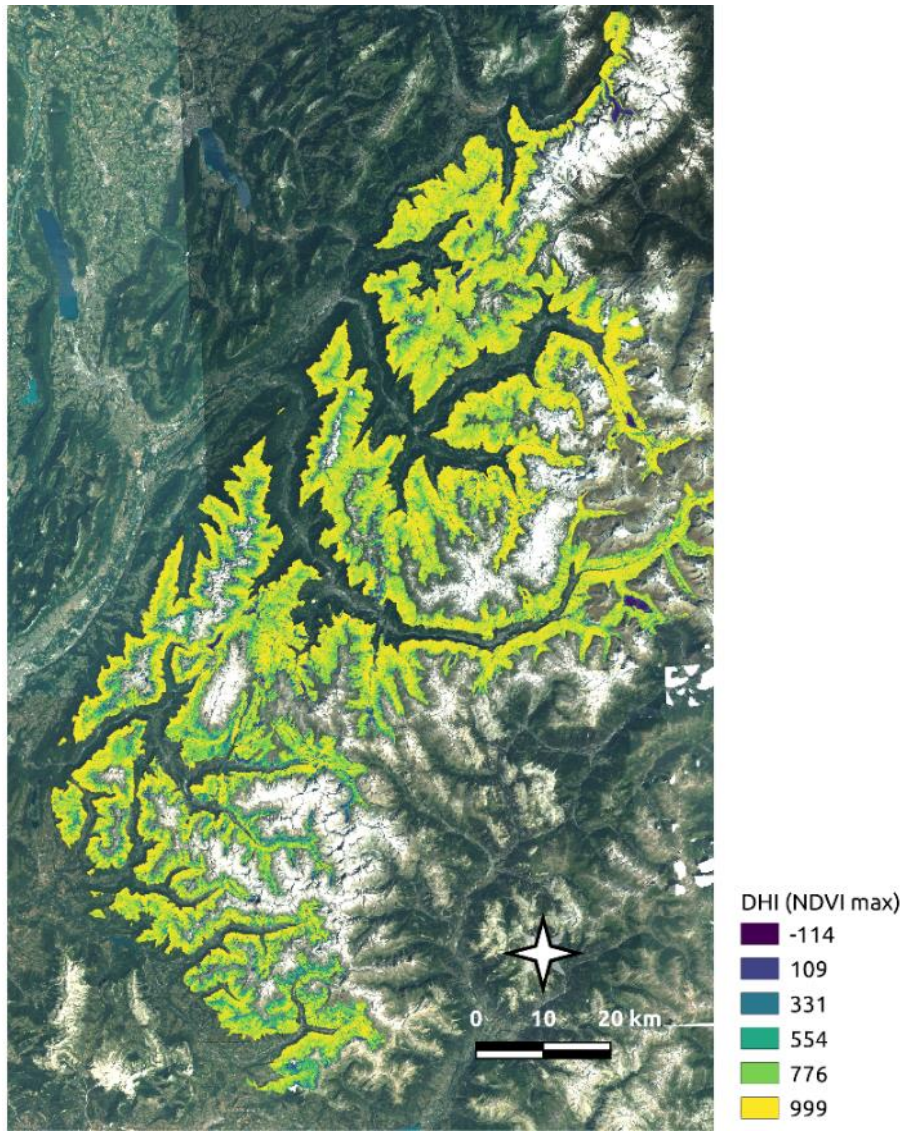

***Annexe 3.*** *Distribution du NDVI Max (DHI) à l'intérieur de la RBC3. Le fond de carte correspond à la mosaïque d'images SPOT6-7 « pansharpened » (multispectrale).*

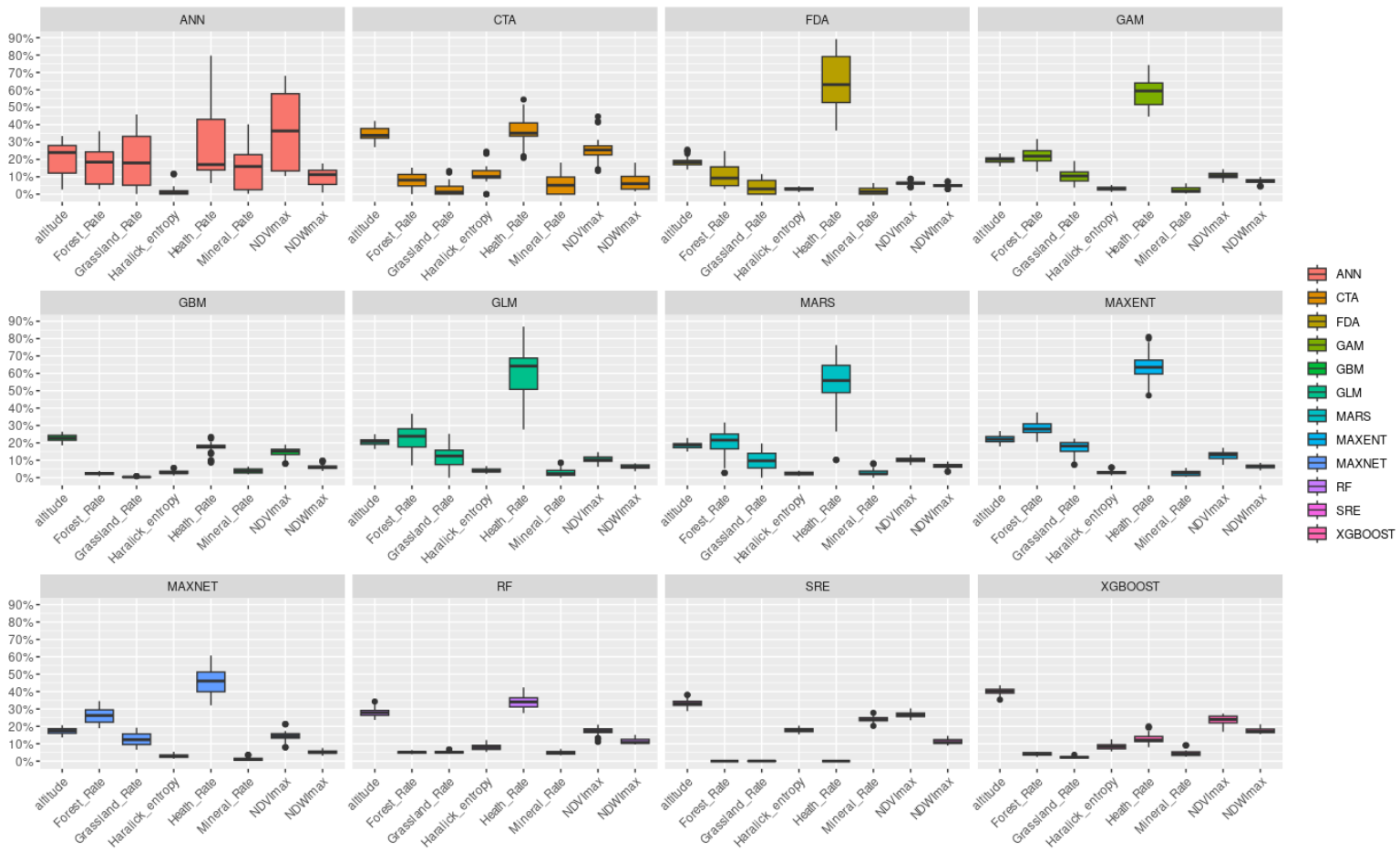

***Annexe 4.*** *Importance relative des variables environnementales des différents algorithmes fournis par* biomod2 *(modèles individuels).*

# X) Bibliographie